\documentstyle[11pt]{article}

\newcommand{\BE}{\begin{equation}}
\newcommand{\EE}{\end{equation}}
\newcommand{\BA}{\begin{eqnarray}}
\newcommand{\EA}{\end{eqnarray}}

\begin{document}
\begin{titlepage}

\vspace*{1mm}
\begin{center}

            {\LARGE{\bf On the infrared behaviour \\ 
                 of the (singlet) Higgs propagator  }}

\vspace*{14mm}
{\Large  M. Consoli }
\vspace*{4mm}\\
{\large
Istituto Nazionale di Fisica Nucleare, Sezione di Catania \\
Corso Italia 57, 95129 Catania, Italy}
\end{center}
\begin{center}
{\bf Abstract}
\end{center}

We present a simple semi-perturbative
argument in favour of a peculiar infrared
behaviour of the (singlet) Higgs propagator. On the basis of `triviality' 
one expects a continuum limit with a
two-point function $\Gamma_2(q) \to (q^2 + M^2_h)$. However, this  is not
valid in the limit $q \to 0$ where one actually  finds a singular 
behaviour. This 
is in agreement with both non-perturbative analyses of the effective potential
and with lattice computations of the propagator and of the 
zero-momentum susceptibility 
in the broken phase. The singular behaviour persists 
in an O(N) continuous-symmetry theory, the case
first pointed out by Symanzik, and
supports the existence of an extremely weak $1/r$ potential
that does not disappear when coupling the scalar fields to
gauge bosons. 
\end{titlepage}

\setcounter{equation}{0}
\section{Introduction}

\par
The generally accepted `triviality' \cite{book} of 
$\lambda\Phi^4$ theories in four space-time dimensions is usually
interpreted within leading-order
perturbation theory in a very intuitive way. One starts with the perturbative
one-loop $\beta-$function
\BE
    \beta_{\rm pert}(\lambda)= {{3\lambda^2}\over{16\pi^2}} + 
{\cal O}(\lambda^3)
\EE
   and integrates the differential equation
\BE 
         {{dx}\over{x}}= {{d\lambda}\over{\beta(\lambda)}}
\EE
between a fixed energy scale $x=\mu$ and the `Landau pole' $x=\Lambda$ where
$\lambda(x)=+\infty$. In this way, at energy scales                  
${\cal O}(\mu)$, the theory is governed by a
1PI 4-point function 
\BE
\lambda \equiv \lambda (\mu)= 
{{32 \pi^2 }\over{3 \ln{{\Lambda^2}\over{\mu^2}} }}
\EE
that would vanish in the
continuum limit where the ultraviolet cutoff $\Lambda \to \infty$ within 
all loop diagrams. In this 
picture, differently from the original renormalization-group approach where
$\lambda(\mu)$ is changed along a given integral curve of Eq.(1.2) by changing 
$\mu$, one
considers all possible integral curves at the same time. After that, one
changes $\lambda$, at fixed $\mu$, depending on the magnitude of the
Landau pole associated with the various integral curves. In this way, 
the `bare' theory is always defined in the infinite-coupling limit, i.e. as 
in an Ising model. However, at any finite scale $\mu$, 
the continuum limit corresponds to a vanishingly small
interaction strength so that perturbation theory in the 
small parameter $\lambda(\mu)$ should provide
very accurate predictions for low-energy physical observables.

By adopting this view of `triviality' in the
spontaneously broken phase of $(\lambda\Phi^4)_4$, 
the euclidean propagator for the Higgs particle should approach the form
\BE
\tilde G(q) \to {{Z_{\rm prop} }\over{q^2 + M^2_h}} 
\EE
with a residue
\BE
Z_{\rm prop} = 1 - |{\cal O} (\lambda)| \to 1
\EE
consistent with the
K\"allen-Lehmann decomposition that dictates the
spectral function $\rho_h(s)$ to approach $\delta(s-M^2_h)$ in the continuum limit
of a `trivial' theory. 

\par
This simple picture neglects, however, that
the origin of spontaneous symmetry breaking 
is not necessarily of perturbative nature.
Indeed,  one may be faced with
non-analytic $1/\lambda$ effects 
so that, even with an infinitesimal two-body interaction strength, there may be
non-perturbative effects. To understand the possible implications, 
one should remember 
the case of superconductivity. This is due to the basic instability of a
normal Fermi system in the presence of an {\it infinitesimally small}
attractive interaction between electrons. 
Due to the presence of a very large density of quantum states near the Fermi 
surface, superconductivity is a non-perturbative phenomenon. 

In the case of spontaneous symmetry breaking with an elementary scalar field
the delicate issue concerns the limit $q \to 0$ that requires some
care in the case of a macroscopic occupation of the same quantum state, i.e.
of Bose-Einstein condensation. 
Just for this reason, and contrary to the most naive expectations,
the approach to the continuum limit in the spontaneously broken phase of
$(\lambda\Phi^4)_4$ theories may contain unexpected features. These are 
discovered
whenever one takes seriously `triviality' as a technical statement 
controlling the approach to the continuum theory and, thus, supporting the 
idea of a trivially free fluctuation field with a gaussian quantum measure 
for $\Lambda \to \infty$.  In fact, as pointed out
in refs.\cite{new, zeit, rit2, agodi}, assuming gaussian 
(and post-gaussian) ansatz for the
ground state wave functional, one finds
\BE
{{\Gamma_2(0)}\over{M^2_h}}= 
{\cal O}(1/\ln \Lambda) \to 0
\EE
once the zero-momentum two-point function is computed through the effective
potential
\BE
\Gamma_2(0)\equiv  
       \left. \frac{ d^2 V_{\rm eff}}{d \phi^2} \right|_{\phi=v} 
\EE 
As pointed out in refs.\cite{new,zeit,rit2,agodi}, Eq. (1.6)
requires a non-trivial re-scaling 
\BE
 Z\equiv Z_\phi={\cal O}(\ln \Lambda) \to \infty
\EE
 of the vacuum field $\phi$
\BE
\phi^2_R \equiv {{\phi^2 }\over{Z_\phi}}          
\EE
in order to match the quadratic shape of the effective potential with the
physical mass $M_h$ defined from the $q \neq 0$ behaviour of the propagator
\BE
\left. \frac{ d^2 V_{\rm eff}}{d \phi_R^2} \right|_{\phi_R=v_R} \equiv M_h^2.
\EE
As such, $Z=Z_\phi$ is quite distinct from the `trivial' re-scaling  
$Z=Z_{\rm prop}$ in Eq.(1.5), and one may 
obtain a continuum limit where, although $M_h$ vanishes in units of the 
bare $v$, {\it both} $M_h$ and $v_R$ are finite 
quantities (with potentially 
important implications for the commonly quoted upper bounds 
on the Higgs mass from `triviality' \cite{lang}).

The result in Eq.(1.6) is also striking for the following reason. 
The euclidean value $q=0$ corresponds, indeed, to
the single point $(q_o,{\bf{q}})=0$ in the
continuum theory. However, in the cutoff theory, Eq.(1.6) implies that
there is a {\it region} of 3-momenta say
${\bf{q}}^2 \ll {\cal O}(1/\ln \Lambda) M^2_h$
where the energy spectrum is {\it not}
$\tilde{E}( {\bf{q}}) = \sqrt{ {\bf{q}}^2 + M^2_h}$.
This region, although infinitesimal 
in units of $M_h$, can have a physical meaning
(for $M_h = {\cal O}(10^2)$ GeV, think of the values 
$|{\bf{q}}|\ll 10^{-5}$ eV/c corresponding to wavelengths much larger than
1 cm).

The previous result was obtained in the formalism of the gaussian ( and 
post-gaussian) approximations to the effective potential.
In the next section, we shall outline a simple semi-perturbative 
argument that leads to the same conclusions and can help to understand in 
a more intuitive way the singular nature of the limit
$q \to 0$ when approaching the continuum theory from the broken phase.
\setcounter{equation}{0}
\section{A simple semi-perturbative calculation}
Let us consider a one-component $\lambda\Phi^4$ theory
\BE
         {\cal L} = {{1}\over{2}} (\partial \Phi)^2 - U(\Phi)
\EE
with a classical potential ($\lambda>0$)
\BE
U(\Phi)={{1}\over{2}} m^2_B\Phi^2+ {{\lambda}\over{4!}}\Phi^4
\EE
Classically, non-vanishing 
constant field configurations 
$\phi=\pm v$ occur where $U'(\pm v)=0$. At these values, one gets a quadratic
shape
\BE  
          U''(\pm v)= {{\lambda v^2}\over{3}} \equiv M^2_h
\EE
that represents the well known classical result for the Higgs mass.
\par 
In the quantum theory, the question of vacuum stability is more subtle and
one has to replace $U(\phi)$ with 
the quantum effective potential $V_{\rm eff}(\phi)$.
However, the basic expectation is  
that the excitation spectrum of the broken phase will
maintain the Lorentz-covariant form
$\tilde{E}( {\bf{q}}) = \sqrt{ {\bf{q}}^2 + M^2_h}$ down to ${\bf{q}}=0$ so that
$M_h$ should coincide with $\tilde{E}(0)$ that represents the energy-gap of the
broken phase.
\par 
To check this prediction, let us
consider the one-loop structure of the gap-equation for the 
euclidean  two-point function
\BE
\tilde G^{-1}(q)\equiv\Gamma_2(q) =q^2 + m^2_B+ {{\lambda\phi^2}\over{2}} 
+{{\lambda}\over{2}}\int {{d^4 k}\over{(2\pi)^4}}\tilde G(k)
 -{{\lambda^2\phi^2}\over{2}}
\int {{d^4 k}\over{(2\pi)^4}}\tilde G(k)\tilde G(k+q) 
\EE 
together with 
the vanishing of the one-loop tadpoles, i.e.
\BE
T(\phi)\equiv m^2_B + {{\lambda\phi^2}\over{6}} 
+{{\lambda}\over{2}}\int {{d^4 k}\over{(2\pi)^4}}\tilde G(k)=0
\EE
Eqs.(2.4) and (2.5)
can be understood, for instance, as coupled minimization equations of the
effective potential for composite operators introduced by Cornwall, Jackiw and
Tomboulis \cite{cjt}. As such, they
are non-perturbative, being equivalent to the {\it all-order} resummation of 
one-loop graphs with the tree-level propagator in the external 
background field $\phi$
\BE
 \tilde{G}(q)_{\rm tree} ={{1}\over{q^2 + m^2_B+ {{\lambda\phi^2}\over{2}} }}
\EE
Notice that for $\phi \neq 0$, the gap-equation would also
contain one-particle reducible terms (i.e. not contained in $\Gamma_2(q)$)
proportional to the zero-momentum limit of the shifted field propagator
$\tilde{G}(0)$.  In this sense, by considering the 1PI 
gap-equation, we are {\it assuming} a non-singular  zero-momentum limit,  
even for $ \phi = \pm v$ where $\pm v$ are the solutions of (2.5) and represent, 
to this order, the minima of the effective potential.

The possibility to solve simultaneously Eqs.(2.4) and (2.5), in the
continuum limit of the regularized theory by using Eqs. (1.3) - (1.5), amounts
to describe spontaneous symmetry breaking as a {\it quantum} phenomenon of
vacuum instability consistently with (the intuitive interpretation of) 
rigorous quantum field theoretical results on $(\lambda\Phi^4)_4$ theories.

By using (2.5) in (2.4) we get 
\BE
\tilde G^{-1}(q)\equiv\Gamma_2(q)= q^2 +
 {{\lambda v^2}\over{3}} A(q) 
\EE
where 
\BE
A(q)= 1 -{{3\lambda}\over{2}}
\int {{d^4 k}\over{(2\pi)^4}}\tilde G(k)\tilde G(k+q) 
\EE
\par

Notice that a form of the propagator 
as in Eq.(1.4) can be a solution of Eq.(2.4)
only if the coupling $\lambda$ 
is understood as an infinitesimally small quantity, i.e. as in Eq.(1.3) 
with $\mu \sim M_h$. In this case, 
for $G(q) \sim {{1}\over{q^2}}$ at large euclidean $q^2$, one finds
$A(q)=A(0)+ {\cal O}(1/\ln \Lambda)$ and one indeed 
gets a form $\Gamma_2(q)= q^2 + const$, 
up to terms vanishing in the continuum limit 
where $\lambda \to 0$. 
In this sense, renormalized
perturbation theory should be considered  an external input
whose overall consistency with the all-order Eqs.(2.4) and (2.5) 
can only be checked {\it a posteriori}.

\par
By using Eq. (1.4) in
Eqs.(2.7) and (2.8) we obtain the leading-order expression
\BE
        Z_{\rm prop}= {{1}\over
{1 + {{\lambda v^2}\over{3M^2_h }} {{\lambda Z^2_{\rm prop} }\over{64\pi^2}} }}
\EE
and the estimate of the Higgs mass 
\BE
M^2_h=
{{\lambda v^2 Z_{\rm prop}}
\over{3}} ( 1- {{3\lambda Z^2_{\rm prop} }\over{32 \pi^2}}
 \ln{{\Lambda^2}\over{M^2_h}}) 
\EE
At this point we find a strong contradiction. 
Indeed, by using Eq. (1.3) and assuming (1.5) 
we obtain from (2.10)
\BE
          M^2_h = {\cal O} (\lambda^2 v^2)
\EE 
that when inserted in Eq.(2.9), 
does {\it not} produce a
$Z_{\rm prop} \to 1$  when $\Lambda \to \infty$.

It is true that in a more conventional fixed-order 
perturbative calculation one would tend to regard 
(2.10) as the first two
terms in the expansion of the renormalized coupling constant at a scale $M_h$ 
(i.e. the renormalization-group improved 
version of the classical result (2.3)). However, this cannot work in our case
for two reasons.
First, as anticipated, Eqs.(2.4) and (2.5) resum one-loop 
graphs with the tree propagator (2.6) to all orders, so that our results 
cannot be considered {\it fixed order} calculations. 
Second, using the simple form of the propagator (1.4) to solve
Eq.(2.4) requires an infinitesimal $\lambda$, as in (1.3). 
However, if one tries to improve
on (2.4) and (2.5), by
introducing {\it genuine} two-loop terms, the
intuitive interpretation of `triviality' based 
on Eqs.(1.1)-(1.5) is destroyed. This is
due to the presence of a (spurious) ultraviolet fixed point at finite coupling
in the two-loop 
perturbative $\beta-$ function \cite{trivpert} so that, beyond a leading-order
calculation, there are no reasons
for the 1PI four point function $\lambda(\mu)$ to vanish in the continuum
limit as in Eq.(1.3) and for Eqs. (1.4) and (1.5) to be valid. 

The previous results suggest that, contrary to our assumption, either 
`triviality'
cannot be understood within leading-order perturbation theory, or
one is faced with
a singular $\tilde{G}(0)$ when $\phi=\pm v$. Namely, a result
$M^2_h={\cal O}(\lambda v^2)$ from Eqs. (2.5) and (2.4), in order Eq.(2.9) 
to agree with (1.5), 
can only be obtained if there are one-particle reducible contributions 
to the propagator. These, at a generic value of $\phi$, 
are proportional to the combination
\BE
            R= T(\phi) \tilde{G}(0)
\EE
and the case $\phi = \pm v$ should be understood
as a limiting procedure due to a 
possible divergence in $\tilde{G}(0)$. To this end, let us introduce
$\phi^2 \equiv v^2(1+\delta)$ with $|\delta| \ll 1$, 
and define
\BE
g \equiv \tilde{G}(0) {{\lambda v^2}\over{3}}
\EE
Now a non vanishing $R$ requires
\BE
      g \sim  {{1}\over{\delta} }         
\EE
implying that, at $\phi=\pm v$ where $\delta=0$, 
there is a mode whose energy 
vanishes for ${\bf {q}}\to 0$. For this reason, 
the energy-gap of the broken phase, obtained from $\tilde{E}({\bf{q}})$ for 
${\bf{q}} \to 0$, cannot be $M_h$. 

\setcounter{equation}{0}
\section{The zero-momentum discontinuity}
\par

We believe that the discrepancy we have pointed 
out, is related to 
the subtleties of Bose-Einstein condensation \cite{plb} in an {\it almost} ideal gas
\cite{mech} where there is a macroscopic occupation of the same quantum
state. This phenomenon cannot be {\it fully}
understood in a purely perturbative manner. 
However, as anticipated in the Introduction, 
the peculiarity of the zero-momentum limit
is found in other approaches such as:
\par~~~~~~i) variational evaluations of the effective potential
\cite{new,zeit,rit2,agodi} 
\par~~~~~~ii) lattice computations of the propagator and 
of the zero-momentum susceptibility $\chi^{-1}\equiv \Gamma_2(0)$ in the broken
phase \cite{cea}. 
These show that Eq.(1.4), although valid at higher momenta, has 
{\it non-perturbative} corrections 
for $q \to 0$. Indeed, 
these become larger and larger by approaching the continuum limit and therefore
cannot represent perturbative ${\cal O}(\lambda)$ effects
\par
Due to this qualitative agreement with other 
approaches, the semi-perturbative
 calculation of sect.2 can be considered a reasonably 
self-contained treatment of the basic zero-momentum discontinuity.
\par 
The discrepancy 
persists in the case of an O(N) continuous-symmetry 
$\lambda\Phi^4$ theory. To this end, one can check with
a non-perturbative Gaussian effective-action approach \cite{oko}. 
In this case, the minimization conditions 
of the gaussian 
effective potential provide suitable relations that replace
 the `triviality pattern' Eqs.(1.3) and (1.5) and where the equivalent of
$T(\phi)$ plays the role of a mass term for the Goldstone bosons.
By neglecting one-particle reducible contributions in the $\sigma-$field
propagator, one finds
(N-1) massless fields with propagator $D_{\pi}(q)=1/q^2$
and a $\sigma-$field two-point function
of the type as in Eqs.(2.7) and (2.8).  By following the steps
 of ref.\cite{oko}, it is not difficult to check that one gets in this way
the same discrepancy as in sect.2 for the mass of the 
$\sigma-$field. This can also be understood since, for
the Goldstone bosons, the non-perturbative wave functional
of ref.\cite{oko} reproduces `triviality' and
is {\it exact}. In fact, it yields (N-1) non-interacting fields that
decouple from each other and from
the $\sigma-$field. As a consequence, one gets effectively
the same type of $\sigma-\sigma$ interactions
as in our discrete-symmetry case.

\par
On the other hand, in the case of an O(N) continuous symmetry,  
the singular nature of the zero-momentum limit
of the singlet-Higgs propagator is well known. It 
was first pointed out by Symanzik for the linear $\sigma$-model \cite{syma}, 
and later on by Patashinsky and Pokrowsky \cite{pata} and
Anishetty et al. \cite{ani}. 
\par
Symanzik's analysis for the $\sigma-$ field, although purely
perturbative, 
displays the essential features of the phenomenon, i.e. the perturbative 
contradiction between finite 1PI vertex diagrams and finite Green's functions
that introduces the zero-momentum discontinuity.
Just for this reason, he introduced two
different notations, namely
$\Gamma_{\sigma}(0)\equiv M^2$ and 
$\Gamma_{\sigma}(q^2)\equiv (q^2+\bar{M}^2)$, to emphasize that the limit
$q \to 0$ is not defined. 
\par From ref.\cite{pata}, on the other hand, one can get a better 
feeling of what is actually going on. In fact, in the case of a spontaneously 
broken O(N) symmetry, the longitudinal susceptibility is found
\cite{pata}
\BE
\chi_{ ||} ({\bf{q}}) 
\sim {{1}\over{|{\bf{q}}|}}~
arctg{{ |{\bf{q}}|}\over{2\kappa}}
\EE
where 
\BE
         \kappa^2 \sim (\phi -v)^2
\EE
Now, for any $\phi \neq v$, the limit ${\bf {q}} \to 0$ yields a finite result.
However, just in the case $\phi=v$, 
$\chi_{ ||} ({\bf{q}})$ becomes singular when ${\bf {q}} \to 0$ as in our case.
\par
Our results show no qualitative difference 
with respect to the continuous-symmetry case of ref.\cite{oko} and, therefore,
the agreement is not surprising.
Indeed, the zero-momentum discontinuity does {\it not}
depend on the existence of a continuous symmetry of the Lagrangian. Rather, 
its physical origin has to be searched in the presence
of the scalar condensate, i.e. in
the phenomenon of Bose-Einstein condensation that
leads to a gap-less mode and to a long-range $1/r$ potential 
\cite{gravity,ferrer}. 

This can be understood as follows. 
As discussed in refs.\cite{return,ritschel,zeit,mech}, variational 
approximations to $V_{\rm eff}$ describe spontaneous symmetry 
breaking as an {\it infinitesimally} weak first-order transition. This occurs
when the mass-gap at $\phi=0$, say $0\leq m^2 \leq m^2_c$, is still 
positive, but in a `hierarchical' relation 
\BE
             {{m^2}\over{M^2_h}}= {\cal O}({{1}\over{
\ln{{\Lambda^2}\over{\mu^2}} }}) 
\EE
with the mass scale of the broken phase \cite{hiera}.

As discussed in ref.\cite{gravity}, 
the gap-less mode with $\tilde{E}({\bf{q}})= const.|{\bf{q}}|$ is associated 
with the infinitesimal region of momenta
\BE
            {\bf{q}}^2 \ll {{M^2_h}\over 
{\ln{{\Lambda^2}\over{\mu^2}} }}  
\EE
that goes, indeed, into the single point $(q_o,{\bf{q}})=0$ in the
continuum limit $\Lambda \to \infty$.
However, in the cutoff theory this region
defines the {\it non-relativistic} limit
$|{\bf {q}}| \ll m$ where $m$ is the mass of the quanta in the condensate
(see Eq. (3.3)).

In this regime, any scalar condensate, whatever its origin
may be, is a highly correlated structure with long-range order 
due to the coherence effects associated with the phase of the
non-relativistic condensate wave-function \cite{anderson,trento}.
Therefore, for ${\bf{q}} \to 0$, 
the deviations from a Lorentz-covariant energy
spectrum $\tilde{E}({\bf {q}})= \sqrt{ {\bf {q}}^2 + M^2_h }$ are not 
surprising. 

Now, for $M_h = {\cal O}(10^2)$ GeV, 
it is a matter of taste to decide whether, for instance, 
values $|{\bf{q}}|\ll 10^{-5}$ eV/c 
(corresponding to wavelengths much larger than 1 cm) may be considered 
infinitesimal or not. In the case of a positive answer, 
these deviations from exact Lorentz-covariance on such scales
should be taken seriously.
Indeed, as discussed in \cite{gravity}, 
the associated extremely weak $1/r$ potential does not
disappear when coupling the scalar fields to gauge bosons.

\end{document}